\documentclass[a4paper,11pt,amssymb,eqsecnum]{article}
\pdfoutput=1
\usepackage{jcappub}
\usepackage{graphicx}
\usepackage{bm}
\usepackage{amsmath}
\usepackage{hyperref}
\usepackage[T1]{fontenc}

\def \lleq {\lower0.9ex\hbox{ $\buildrel < \over \sim$} ~}
\def \ggeq {\lower0.9ex\hbox{ $\buildrel > \over \sim$} ~}

\def \beq  {\begin{equation}}
\def \eeq  {\end{equation}}
\def \ber  {\begin{eqnarray}}
\def \eer  {\end{eqnarray}}

\newcommand{\newc}{\newcommand}
\newc{\be}{\begin{equation}}
\newc{\ee}{\end{equation}}
\newc{\ba}{\begin{eqnarray}}
\newc{\ea}{\end{eqnarray}}
\newc{\bea}{\begin{eqnarray*}}
\newc{\eea}{\end{eqnarray*}}
\newc{\D}{\partial}
\newc{\ie}{{\it i.e.} }
\newc{\eg}{{\it e.g.} }
\newc{\etc}{{\it etc.} }
\newc{\etal}{{\it et al.}}
\newcommand{\nn}{\nonumber}
\newc{\ra}{\rightarrow}
\newc{\lra}{\leftrightarrow}
\newc{\lsim}{\buildrel{<}\over{\sim}}
\newc{\gsim}{\buildrel{>}\over{\sim}}

\title{Gravitational wave source counts at high redshift and in models with extra dimensions}

\author{Juan Garc\'ia-Bellido,}
\author{Savvas Nesseris}
\author{and Manuel Trashorras}
\affiliation{Instituto de F\'isica Te\'orica UAM-CSIC, Universidad Auton\'oma de Madrid,
Cantoblanco, 28049 Madrid, Spain}
\emailAdd{juan.garciabellido@uam.es}
\emailAdd{savvas.nesseris@csic.es}
\emailAdd{manuel.trashorras@csic.es}

\abstract{Gravitational wave (GW) source counts have been recently shown to be able to test how gravitational radiation propagates with the distance from the source. Here, we extend this formalism to cosmological scales, i.e. the high redshift regime, and we discuss the complications of applying this methodology to high redshift sources. We also allow for models with compactified extra dimensions like in the Kaluza-Klein model. Furthermore, we also consider the case of intermediate redshifts, i.e. $0<z\lesssim1$, where we show it is possible to find an analytical approximation for the source counts $\frac{dN}{d(S/N)}$. This can be done in terms of cosmological parameters, such as the matter density $\Omega_{m,0}$ of the cosmological constant model or the cosmographic parameters for a general dark energy model. Our analysis is as general as possible, but it depends on two important factors: a source model for the black hole binary mergers and the   GW source to galaxy bias. This methodology also allows us to obtain the higher order corrections of the source counts in terms of the signal-to-noise $S/N$. We then forecast the sensitivity of future observations in constraining GW physics but also the underlying cosmology by simulating sources distributed over a finite range of signal-to-noise with a number of sources ranging from 10 to 500 sources as expected from future detectors. We find that with 500 events it will be possible to provide constraints on the matter density parameter at present $\Omega_{m,0}$ on the order of a few percent and with the precision growing fast with the number of events. In the case of extra dimensions we find that depending on the degeneracies of the model, with 500 events it may be possible to provide stringent limits on the existence of the extra dimensions if the aforementioned degeneracies can be broken.}

\begin{document}
\maketitle
\flushbottom

\section{Introduction}

The recent first detection by LIGO of gravitational waves (GWs), emitted from a merging black hole (BH) binary system, offers a new opportunity to understand gravity in the strong field regime but also observe the evolution of the universe in a completely new way \cite{Abbott:2016blz}. Just as we can extract cosmological information from standard sirens, we will now be able to do the same with gravitational wave source count distributions.

For gravitational waves emitted from an inspiralling black hole binary system at high redshifts we have that the amplitude of the waves (or strain) $h_{+\times}(t)$ can be given in terms of the frequency of the waves $f$, the chirp mass $M_c$, the cosmological redshift $z$ of the objects and their luminosity distance $D_L(z)$ \cite{Deffayet:2007kf}:
\be
h_{+\times}(t) \propto \frac{\left((1+z) M_c\right)^{5/3}f_{obs}^{2/3}}{D_L(z)},
\ee
where the two transverse GW polarizations are represented by the $+\times$ and the luminosity distance is given by\footnote{Note that in later sections we will adopt natural units with $c=8\pi G=1$.}
\be
D_L(z)=\frac{c}{H_0}(1+z)\frac{1}{\sqrt{-\Omega_K}}\sin\left(\sqrt{-\Omega_K} \int_0^z\frac{dz'}{H(z')/H_0}\right),
\ee
where $c$ is the speed of light in vacuum, $H_0$ is the Hubble constant, $\Omega_K=-\frac{K}{a_0^2 H_0^2}$ is the dimensionless curvature density parameter, $K=(-1,0,1)$ and $a_0$ is the present value of the scale factor  \cite{Weinberg:2008zzc}.

In order to extract cosmological information we need to know the propagation of GWs in space, from the source to the detector, and its effect on the detector in terms of the signal-to-noise ($S/N$) ratios, which is given by:
\ba
|\tilde{h}(f)|^2&=& H_0^2 \Omega_{GW}(f) \frac{3}{8 \pi^2} f^{-4}, \\
(S/N)^2&=& 4 \int_0^\infty \frac{|\tilde{h}(f)|^2}{S_n(f)}df,
\ea
where $S_n(f)$ is the spectral noise density of the detectors, $\Omega_{GW}(f)$ is the dimensionless density parameter for the gravitational waves and $\tilde{h}(f)$ is the  frequency-domain strain \cite{Moore:2014lga}.

Recently, it was shown that, neglecting for the time being the dependence on the chirp mass, the expected low redshift ($z\ll1$) source counts scale as \cite{Calabrese:2016bnu}
\be
\frac{dN}{d(S/N)}\sim (S/N)^{-4}.
\ee
Furthermore, it was also shown that the expected GW source count distributions can place strong constrains on the underlying theory of gravity without knowing the actual redshift of the GW sources. The methodology of Ref.~\cite{Calabrese:2016bnu} does not require or assume any redshift dependence for the GW sources and consequently it can only be valid in short, i.e. not cosmological ($z\ll1$), distances. However, in general we would expect GW sources to be present also in higher redshifts, therefore the differential source counts should pick up relativistic corrections due to the expansion of the universe. In fact, the authors in Ref.~\cite{Belczynski:2016obo} have recently produced updated estimates for the merger rate density of BH-BH binaries as a function of the cosmological redshift $z$. At this point it should be noted that it is unclear whether the hosts of the BH binaries can always be really observed. Progress was recently made in this direction, as it was proposed in Ref.~\cite{Oguri:2016dgk} that the cross-correlation between the spatial distributions of GW sources and galaxies with known redshifts could be used as a new way to constrain the distance redshift relation.

Therefore, anticipating future progress in better determining the redshifts of BH binaries or at least constraining their distance-redshift relation, in this paper we will also  consider the high $z$ limit of the differential GW source counts, by extending the methodology of Ref.~\cite{Calabrese:2016bnu}. In this limit we will assume that either the number density of BH is approximately constant per comoving volume or that it is proportional to that of the galaxies, assuming a GW source to galaxy {\it bias} $b_{BH}$. Since only a given fraction of BH binaries coalesce to emit in the form of GW signal, we can then compute the number counts using the luminosity of these sources as a function of redshift. However, the first case is not very realistic since in Ref.~\cite{Belczynski:2016obo} it was found that the merger rate at small to intermediate redshifts up until $z\lesssim2.5$ increases to $\sim2\cdot10^{-6} \textrm{Mpc}^{-3} \textrm{yr}^{-1}$, but then decreases to a rate of $\sim2\cdot10^{-8} \textrm{Mpc}^{-3} \textrm{yr}^{-1}$ at a redshift of $z\sim14$ and finally goes quickly to zero. In the same reference the authors estimated the merger rate density for BH-BH binaries based on simulations with a total mass for the binary systems of the order of $M_{tot}<240 M_{\odot}$.

In a different vein, we will also consider the case when the underlying gravity theory is different from General Relativity in four dimensions. In particular, we also allow for models with a compactified extra dimension like in the Kaluza-Klein (KK) model, where the propagation of GWs will depend on the radius of the fifth compact dimension or alternatively on the mass of the graviton. Note that the LIGO collaboration has already provided a bound on the graviton mass $m_g < 1.2\times10^{-22}$ eV$/c^2$, see Ref.~\cite{Abbott:2016blz}, which translates into a distance $R_c > 0.3$~pc. Old studies of clusters of galaxies actually put a much stronger constraint $m_g < 10^{-29}$ eV$/c^2$, or $R_c > 10$~Mpc \cite{Goldhaber:1974wg}, while for a comprehensive list see Ref.~\cite{Agashe:2014kda}. Here we consider the compact extra dimension to be larger than about $R_c \simeq 1$ Gpc.

\section{Gravity Wave Source Counts  \label{sec:NC}}
\subsection{The high redshift limit in 4 dimensions \label{sec:highz}}

In this section we calculate the source counts to all cosmological redshifts $z$. The signal-to-noise associated to a GW source at redshift $z$ in 4 space-time dimensions and at a distance $D_L(z)$ is given by \cite{Deffayet:2007kf}:
\be
S/N \propto \frac{(1+z)^{5/3}}{D_L(z)}. \label{eq:SNz}
\ee

\begin{figure}[!t]
\centering
\vspace{0cm}\rotatebox{0}{\vspace{0cm}\hspace{0cm}\resizebox{0.80\textwidth}{!}{\includegraphics{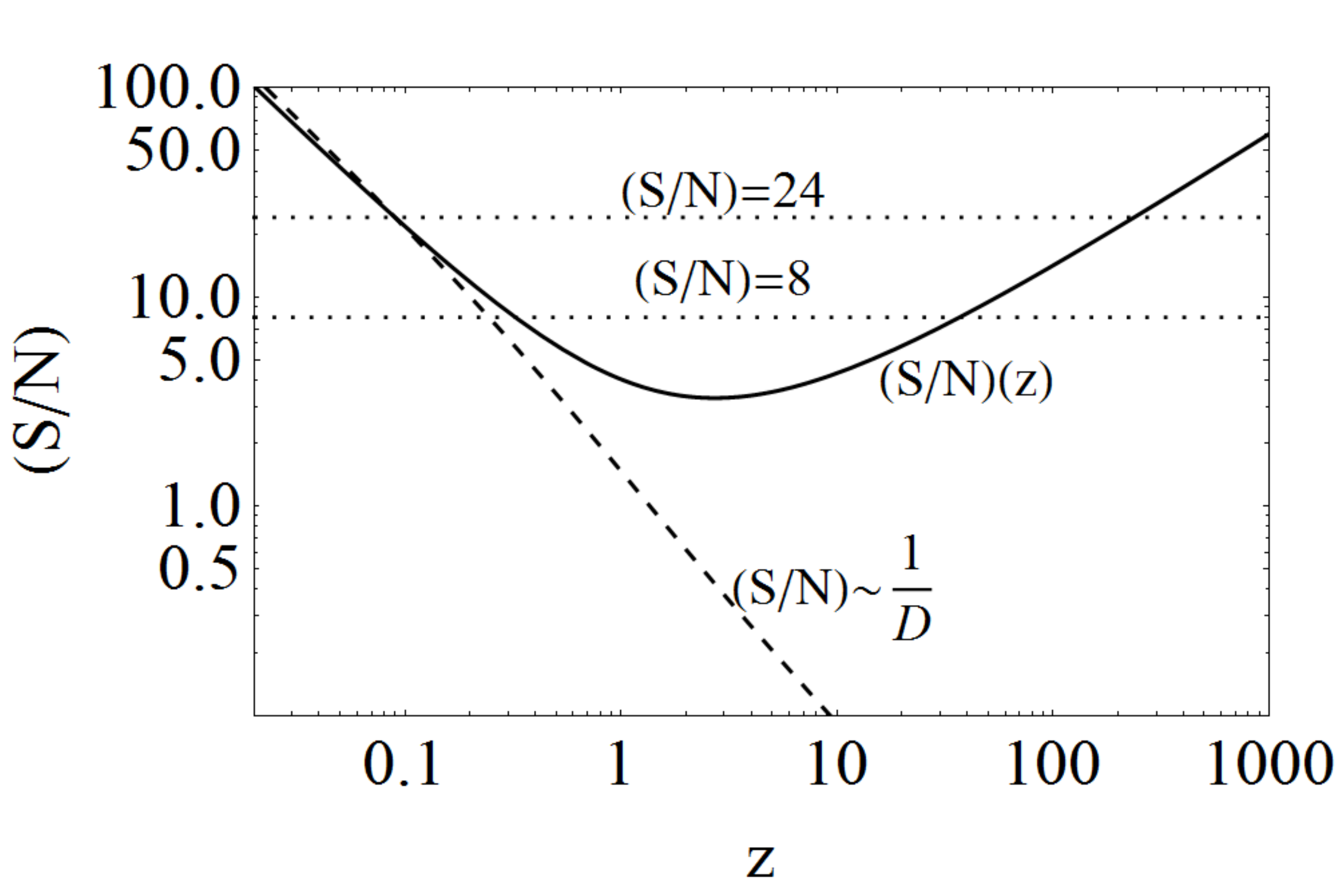}}}
\caption{The signal-to-noise $S/N$ of Eq.~(\ref{eq:SNz}) as a function of redshift for the $\Lambda$CDM model with $\Omega_{m,0}=0.27$ (black solid line) and the low-$z$ prediction of $S/N\propto\frac{1}{D}$ (dashed line). We have normalized the signal-to-noise so that at a redshift of $z_{LIGO}=0.09$, the redshift of the event observed by LIGO, the signal-to-noise corresponds to $S/N=24$, ie $S/N(z_{LIGO})=24$. The dotted lines correspond to the minimum $S/N$ for the LIGO experiment and the value of the observed event. It should be noted that the minimum $S/N$ occurs at $z\sim2.72$ for the $\Lambda$CDM model. \label{fig:SNz1}}
\end{figure}

In Fig.~\ref{fig:SNz1} we show the signal-to-noise $S/N$ of Eq.~(\ref{eq:SNz}) as a function of redshift for the flat $\Lambda$CDM model with a matter density parameter of $\Omega_{m,0}=0.27$ (black solid line) and the low redshift expression of $S/N\propto\frac{1}{D}$ (dashed line) extrapolated to high $z$. We have normalized the signal-to-noise so that at a redshift of $z_{LIGO}=0.09$, the redshift of the event observed by LIGO, the signal-to-noise corresponds to $S/N=24$, ie $S/N(z_{LIGO})=24$. The dotted lines correspond to the minimum $S/N$ for the LIGO experiment and the value of the observed event respectively. It should be noted that the minimum $S/N$ occurs at $z\sim2.72$ for the $\Lambda$CDM model.

As can be seen in Fig.~\ref{fig:SNz1}, compared to the low redshift expression of $S/N\propto\frac{1}{D}$, in reality the signal-to-noise has some interesting features. First of all, due to the expansion of the universe the signal-to-noise has a minimum of $S/N\sim3.29$ at $z\sim2.72$ for the $\Lambda$CDM model, well below the current detector threshold of $(S/N)_{min}=8$. This implies that for a given configuration of the BH-BH binary system that produces the GW event, contrary to the naive $\sim1/D$ behavior, there is a physical limit to the minimum signal-to-noise necessary to observe distant events. However, in general the GW emission and the observed strain will also depend on the properties of the sources and mainly the chirp masses and as a result it would be possible to detect events at the same redshift with different values for the $S/N$.

It should be noted that the difference of the simple law $S/N\propto\frac{1}{D}$ and Eq.~(\ref{eq:SNz}) is quite dramatic even at low redshifts. Specifically it is easy to see that the difference of the two expressions is independent of cosmology and scales as $\textit{\textrm{diff.}} \simeq 1-C/(1+z)^{5.3}$ where $C$ is a constant and it is $10\%$ at $z\sim0.16$, $20\%$ at $z\sim0.25$ and $50\%$ at $z\sim0.65$. This of course implies that neglecting to use the proper expression of Eq.~(\ref{eq:SNz}) could easily add significant errors in the interpretation of the results.

At high redshifts the signal-to-noise counterintuitively grows again due to the expansion of the universe as the $(1+z)^{5/3}$ term overtakes the luminosity distance $D_L(z)$. The physical explanation of this is that since the signal-to-noise is a quantity integrated over all the frequencies $f$, so at high redshifts this creates a volume effect due to the many sources and we now have more events. This is analogous to the well known effect of the maximum of the angular diameter distance at intermediate redshifts, see Ref.~\cite{Krauss:1992sa}.

The derivative of the signal-to-noise with respect to the redshift $z$ is given by:
\be
\frac{d(S/N)}{dz}\propto (1+z)^{2/3}\frac{~5 D_L(z)-3 (1+z) D_L'(z)}{D_L(z)^2 },
\ee
and as can be seen from Fig.~\ref{fig:SNz1}, the distribution shows a minimum at some redshift $z\in[2.1,2.7]$, depending on the cosmological parameters. We will discuss the implications of this in what follows. Then, the differential number counts will be:
\be
\frac{dN}{d(S/N)}=\frac{dN}{dz}/\frac{d(S/N)}{dz},
\ee
while the number of sources with redshift between $z$ and $z + dz$ and apparent luminosity between $\ell$ and $\ell+d\ell$ is $N_{BH}(z,\ell)~dz~d\ell$, where $N_{BH}(z,\ell)$ is the number of BH binaries per proper volume at redshift $z$ with apparent luminosity $\ell$. However, we are interested in the number of sources at a given redshift range $[z,z + dz]$ integrated over all apparent luminosities $\ell$, so we have
\be
dN=dz \int_0^\infty N_{BH}(z,\ell) d\ell.
\ee
We can estimate $N_{BH}(z,\ell)~dz~d\ell$ in terms of the $\tilde{N}_{BH}$, the number of sources per proper volume and absolute luminosity $L$ and $L+dL$ as follows, see Section 1.11 in Ref.~\cite{Weinberg:2008zzc} but also Refs.~\cite{Ellis:1971pg,Ribeiro:2003zu}:
\ba
N_{BH}(z,\ell)~dz~d\ell &=& \tilde{N}_{BH}(z,L)~dV~dL \nn \\
&=& 4 \pi~\tilde{N}_{BH}(z,L)~\frac{a(t)^3 r^2 dr~dL}{\sqrt{1-K r^2}}.~~
\ea
Now we can express the differential $dr$ in terms of $dt$ and finally to $dz$ via \cite{Weinberg:2008zzc}:
\be
\frac{dr}{\sqrt{1-K r^2}}=-\frac{dt}{a(t)}=\frac{dz}{H(z)},
\ee
and we have
\ba
N_{BH}(z,\ell)~dz~d\ell&=& 4 \pi~\tilde{N}_{BH}(z,L)~\frac{r^2(z)}{(1+z)^3 H(z)}dz~dL\nn \\
&=& 4 \pi ~\tilde{N}_{BH}(z,L)~\frac{D_L^2(z)}{(1+z)^5 H(z)}dz~dL ,
\ea
where in the last line we have used the fact that $r(z)\equiv \frac{D_L(z)}{1+z}$ is the comoving distance. Finally, the differential number counts become:
\ba
dN&=&dz \int_0^\infty N_{BH}(z,\ell) d\ell \nn\\
&=&dz \int_0^\infty 4 \pi ~\tilde{N}_{BH}(z,L)~\frac{D_L^2(z)}{(1+z)^5 H(z)}dL \nn\\
&=& dz \left(4 \pi \frac{D_L^2(z)}{(1+z)^5 H(z)}\right) \int_0^\infty ~\tilde{N}_{BH}(z,L)~dL\nn\\
&=&dz \left(4 \pi \frac{D_L^2(z)}{(1+z)^5 H(z)}\right) \Phi(z),
\ea
where $\Phi(z)\equiv \int_0^\infty ~\tilde{N}_{BH}(z,L)~dL$ and the number counts become:
\be
\frac{dN}{dz}=4 \pi \frac{D_L^2(z)}{(1+z)^5 H(z)} \Phi(z). \label{ncountsz}
\ee
Finally, we have:
\ba
\frac{dN}{d(S/N)}&=&\frac{dN}{dz}/\frac{d(S/N)}{dz} \nn \\
&=&\frac{12\pi D_L^4(z)\Phi(z)/A}{(1+z)^{17/3}H(z)(5D_L(z)-3(1+z)D_L'(z))},\nn \\ \label{eq:numbercounts1}
\ea
where $A$ is the proportionality constant in Eq.~(\ref{eq:SNz}). In a flat universe Eq.~(\ref{eq:numbercounts1}) becomes:
\ba
\frac{dN}{d(S/N)}|_{flat}=\frac{12\pi D_L^4(z)\Phi(z)/A}{(1+z)^{17/3}\left(2D_L(z)H(z)-3(1+z)^2\right)}.\nn \\
\ea
Next, we will examine two cases. First, if the GW sources are not evolving in time and second when the GW sources are following the distribution of the galaxies.

\subsubsection{Non-evolving GW sources}

In the non realistic case of non-evolving GW sources, studied here only for completeness and since it is much simpler, we have that the number density $\tilde{N}_{BH}(z,L)$ will be proportional to $(1+z)^3$, ie
\be
\tilde{N}_{BH}(z,L)=\tilde{N}_{BH,0}(L)(1+z)^3
\ee
and from Eq.~(\ref{ncountsz}) we have
\be
\frac{dN}{dz}=4 \pi \frac{D_L^2(z)\tilde{N}_{BH,0}}{(1+z)^2 H(z)}, \label{ncountsz1}
\ee
where $\tilde{N}_{BH,0}\equiv \int_0^\infty ~\tilde{N}_{BH}(L)~dL$
and the expected source counts become
\be
\frac{dN}{d(S/N)}=\frac{\tilde{A} D_L^4(z)}{(1+z)^{8/3} H(z) \left(5 D_L(z)-3 (1+z) D_L'(z)\right)} ,\label{eq:dndsnz1}
\ee
where $\tilde{A}\equiv 12 \pi \tilde{N}_{BH,0}/A$ is a constant. In a flat universe with $\Omega_K=0$, Eq.~(\ref{eq:dndsnz1}) simplifies to:
\be
\frac{dN}{d(S/N)}|_{flat}=\frac{\tilde{A}~D_L^4(z)}{(1+z)^{8/3} \left(2H(z) D_L(z)-3(1+z)^2\right)}.\label{eq:dndsnz2}
\ee

\begin{figure}[!t]
\centering
\vspace{0cm}\rotatebox{0}{\vspace{0cm}\hspace{0cm}\resizebox{0.80\textwidth}{!}{\includegraphics{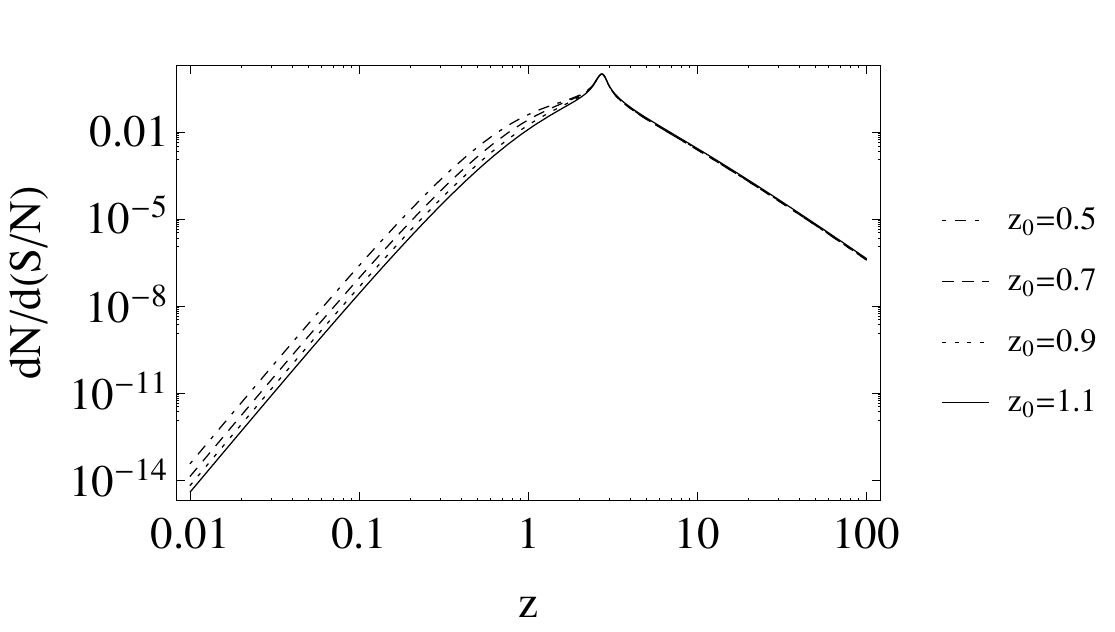}}}
\caption{The regularized differential source counts $\frac{dN}{d(S/N)}$ as a function of redshift, after the singularity has been removed by patching the low $z$ and high $z$ limits with a third order polynomial. We assume a $\Lambda$CDM cosmology with $(\Omega_{m0},w)=(0.27,-1)$.
\label{fig:fixsingularities}}
\end{figure}

\subsubsection{Evolving GW sources}

In practice the GW sources will evolve in time and, to zeroth order, we may assume that their evolution roughly follows the one of their host galaxies. In other words, their number density will be related to the number of sources per proper volume for the host galaxies of the binary black hole system. Most of these black holes are formed at rest and have the velocities of the majority of stars in the galaxy.

Therefore, we assume that the GW sources evolve at a rate proportional to the galaxy number density $\tilde{N}_{BH}(z,L)=b_{BH}(z)~N_{gal}(z,L)$, where $b_{BH}(z)$ is the redshift-dependent BH to galaxy {\it bias} \cite{Oguri:2016dgk}. Thus, we have that:
\be
\frac{dN}{d(S/N)}=\frac{12 \pi}{A} \frac{b_{BH}(z) D_L^4(z)~N_{gal}(z)}{(1+z)^{17/3} H(z) \left(5 D_L(z)-3 (1+z) D_L'(z)\right)},\label{eq:dndsnz3}
\ee
where $N_{gal}(z)=\int_0^\infty N_{gal}(z,L)dL$. Without loss of generality we will assume that $N_{gal}(z)$ is given by the galaxy redshift distribution as measured by several surveys:
\be
\frac{dN_{gal}}{dz}=N_0 \left(\frac{z}{z_0}\right)^2~e^{-\left(\frac{z}{z_0}\right)^{3/2}},
\ee
where $z_0$ is the pivot redshift for the survey, ie $z_0=0.5$ for the Dark Energy Survey~\cite{Sanchez:2010zg} or $z_0=0.7$ for Euclid~\cite{Laureijs:2011gra,Amendola:2012ys} and $N_0$ is the normalization. Then, the number of the galaxies becomes:
\ba
N_{gal}(z)&=&\int_0^z \frac{dN_{gal}}{dz} dz \nn \\
&=&\frac{2N_0 z_0}{3} e^{-(z/z_0)^{3/2}} \left(-(z/z_0)^{3/2}+e^{(z/z_0)^{3/2}}-1\right). \nn \\ \label{eq:ngalz}
\ea
Using Eqs.~(\ref{eq:dndsnz3}) and (\ref{eq:ngalz}) along with an assumption for the bias $b_{BH}(z)$ allows us in principle to estimate the differential source counts $\frac{dN}{d(S/N)}$.

\subsubsection{The number counts as a function of the $S/N$}

By inspecting Eqs.~(\ref{eq:SNz}) and (\ref{eq:dndsnz3}) it is obvious that in order to estimate $\frac{dN}{d(S/N)}$ as a function of the signal-to-noise $S/N$ we need to know both the underlying cosmology in order to calculate the luminosity distance $D_L(z)$, but also the number of sources per proper volume at redshift $z$ given by $\tilde{N}_{BH}(z,L)=b_{BH}(z)~N_{gal}(z,L)$. Therefore, for the sake of simplicity in what follows we will only consider the case of a constant bias $b_{BH}(z)=b_0$.

Unfortunately, at this point we should note that there are two more problems. First, even if we know all of the above information, it is obvious that $\frac{dN}{d(S/N)}$ will only be given as a function of the signal-to-noise $S/N$ parametrically, as eliminating the redshift analytically from Eqs.~(\ref{eq:SNz}) and (\ref{eq:dndsnz3}) is in general not possible, but only numerically or by using a series expansion as we will detail later on. Also, another problem that arises is that as mentioned earlier and illustrated in Fig.~\ref{fig:SNz1} the signal-to-noise as given by Eq.~(\ref{eq:SNz}) has a minimum at $z\sim2$ depending on the cosmological parameters and clearly its first derivative $\frac{d(S/N)}{dz}$ obviously becomes zero at the same $z$. This is obviously problematic as $\frac{d(S/N)}{dz}$ appears in the denominator of the differential number counts $\frac{dN}{d(S/N)}$ as seen in Eq.~(\ref{eq:numbercounts1}), thus leading to a singularity.

\begin{figure*}[!t]
\centering
\vspace{0cm}\rotatebox{0}{\vspace{0cm}\hspace{0cm}\resizebox{0.80\textwidth}{!}{\includegraphics{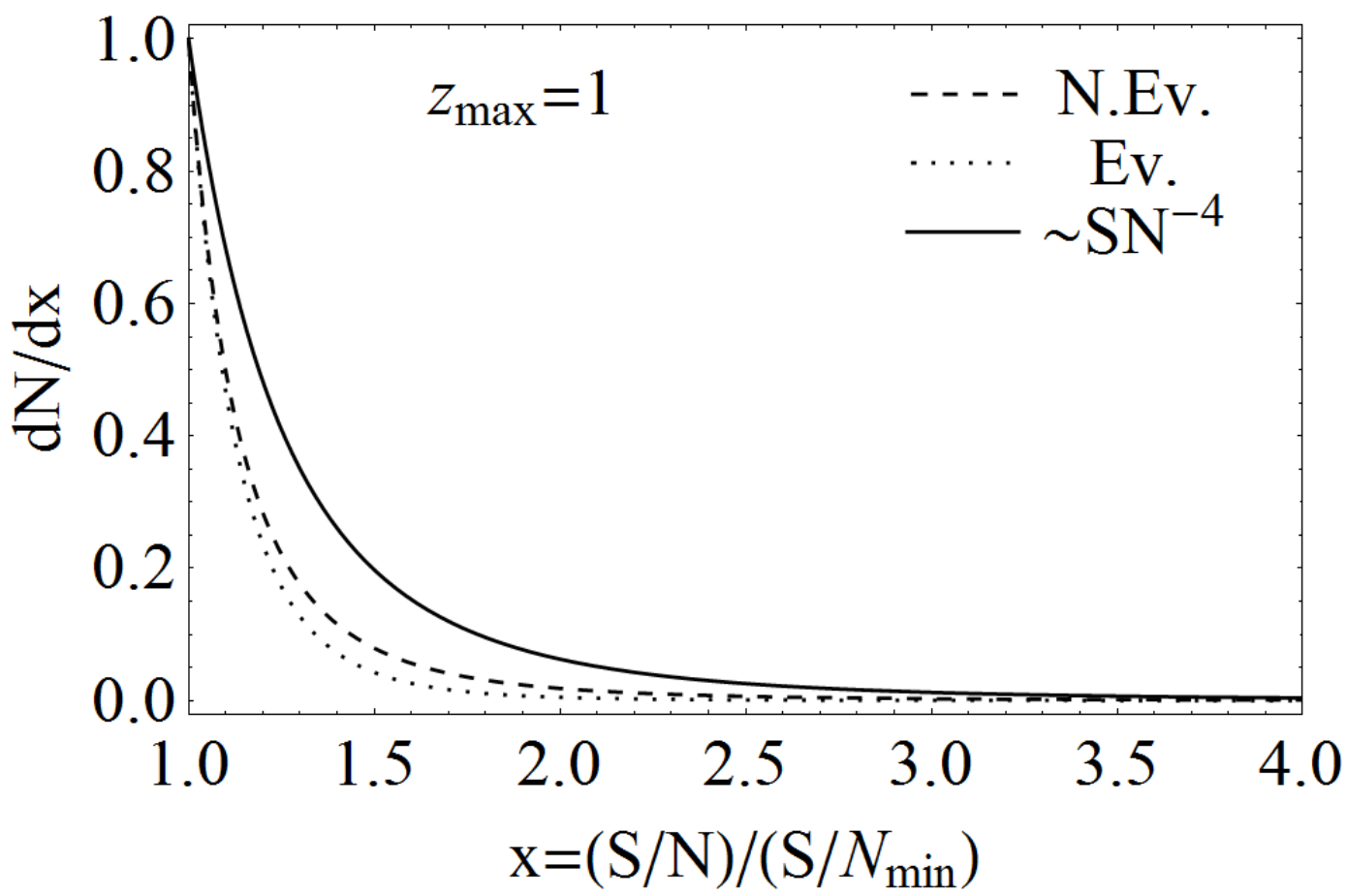}}}
\caption{The differential source counts $\frac{dN}{dx}$ as a function of the normalized signal-to-noise $x=\frac{S/N}{S/N_{min}}$ for $z_{max}=1$. \label{fig:zdependence}}
\end{figure*}

However, it should be noted that Eqs.~(\ref{eq:SNz}) is an approximation and one should not expect it to be valid at all redshifts $z$. Moreover, as already mentioned in Ref. \cite{Sesana:2016ljz} one should not really trust it at $z\simeq2$. Therefore, we will remove the singularity by patching the low $z$ and high $z$ limits with a third order polynomial. The justification for this is that we expect the signal-to-noise of Eqs.~(\ref{eq:SNz}) to be modified by high-order effects, e.g. terms higher than quadrupole in the expansion or even the possibility that the BHs have spins. These effects should in principle induce corrections in the redshift dependence of Eqs.~(\ref{eq:SNz}), thus alleviating the singularity, and for the time being are left for future work. In Fig.~\ref{fig:fixsingularities} we show the regularized differential source counts $\frac{dN}{d(S/N)}$ as a function of redshift for various values of $z_0$, normalized to unity at its maximum value. Now, there is only a small bump where the singularity was and then the source counts decrease once again.

In order to calculate parametrically and numerically the signal-to-noise $S/N$ and the source counts $\frac{dN}{d(S/N)}$ we need to specify the maximum redshift $z_{max}$ up to which we are evolving equations Eqs.~(\ref{eq:SNz}) and (\ref{eq:dndsnz3}). Also, as we will later see in Eq.~\ref{eq:seriesz}, $z$ is directly related to the signal-to-noise $S/N$ and the constant $A$, so choosing to plot the differential source counts only in a specific range of values for the $S/N$, directly implies a related range for $z$.

In Fig.~\ref{fig:zdependence} we show the numerical solutions to the differential source counts $\frac{dN}{dx}$ as a function of the normalized signal-to-noise $x=\frac{S/N}{S/N_{min}}$ for $z_{max}=1$ for the second regularization scheme normalized to one at $x=1$. The black solid line corresponds to the low redshift expectation $dN/d(S/N)\propto (S/N)^{-4}$, the dashed line to the unrealistic but simple case of non-evolving GW sources given by Eq.~(\ref{eq:dndsnz2}) and the dotted line to the one of evolving GW sources given by Eq.~(\ref{eq:dndsnz3}), with a constant bias $b_0$ and $z_0=0.7$. For simplicity and we have chosen a constant bias parameter $b_0=1$.

\begin{figure}[!t]
\centering
\vspace{0cm}\rotatebox{0}{\vspace{0cm}\hspace{0cm}\resizebox{0.80\textwidth}{!}{\includegraphics{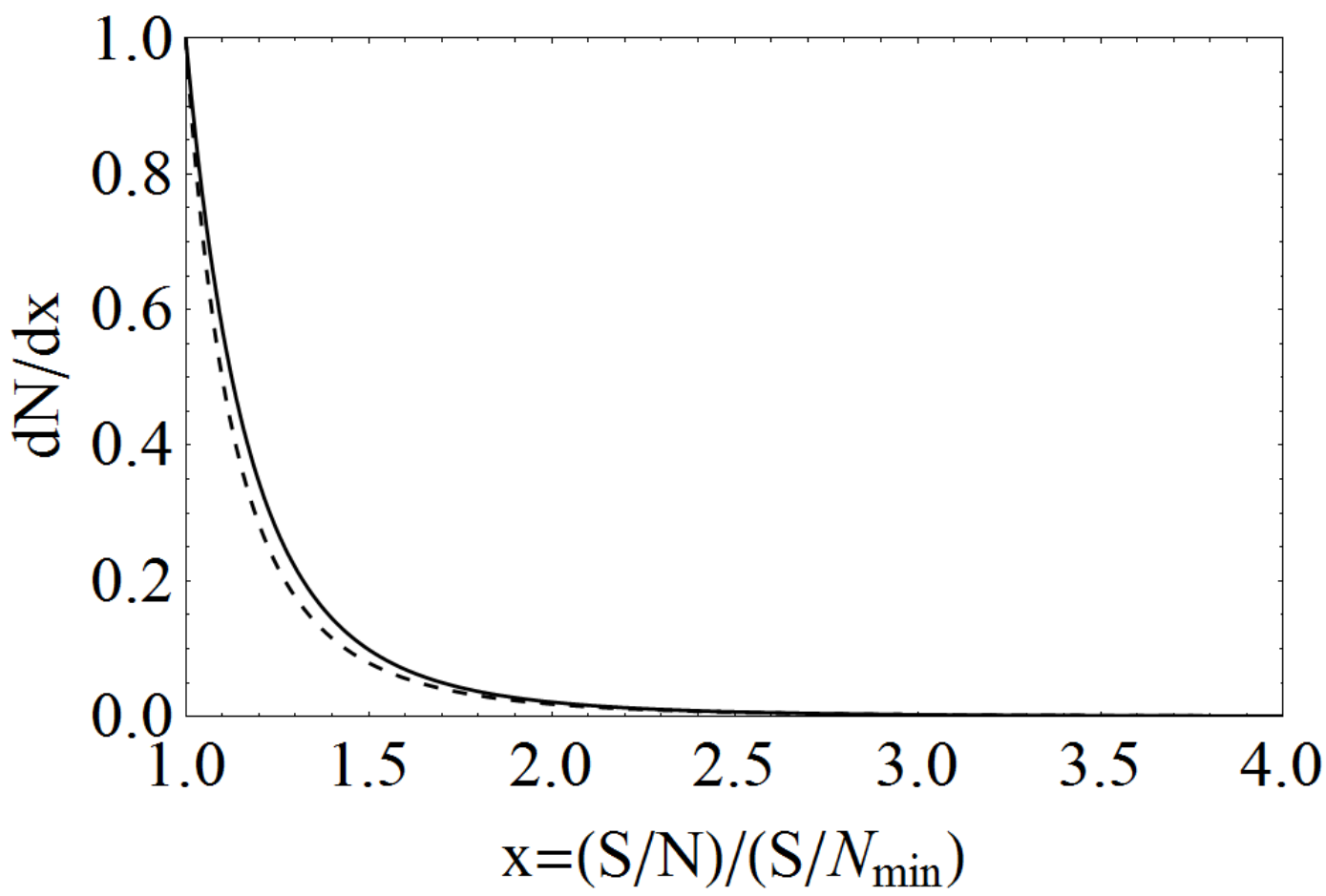}}}
\caption{The differential source counts $\frac{dN}{dx}$ as a function of the normalized signal-to-noise $x=\frac{S/N}{S/N_{min}}$ in the case of non-evolving GW sources for the numerical result (dashed line) and the series expansion of Eq.~(\ref{eq:analz}), which is only valid up to redshifts $z\lesssim1$. For all cases we used $\Omega_{m,0}=0.27$. \label{fig:series}}
\end{figure}

\subsubsection{The intermediate redshift regime and analytical approximations}
As it was made clear in the previous sections, extending our formalism into the high redshift regime is not possible without making certain assumptions like the number density and merger rate density for BH-BH binaries, the distribution of galaxies and finally a regularization scheme in order to remove the singularity that appears due to the presence of the minimum of the $S/N$. This is partly due to the fact that it is not possible to find an analytical expression for the source counts in terms of the signal-to-noise and the fact that in order to give realistic estimates, we require more information about the statistical distribution of the BH binaries. Some progress has been done in this direction, e.g. in Ref.~\cite{Belczynski:2016obo} where the authors provide a realistic estimate for the redshift dependent merger rate density for BH-BH binaries, however a lot remains to be done in the future.

Nonetheless, right now we can at least solve one of the problems, i.e. an analytical expression for the source counts in terms of the signal-to-noise at least in intermediate redshifts. This is possible to do by performing a series expansion in terms of the redshift $z$ and by assuming that the background expansion history of the Universe is given by the cosmological constant model or by a general Dark Energy model described by the {\it deceleration}, {\it jerk} and {\it snap} kinematical quantities.

First, we will study the case of the cosmological constant model. To do so, we perform a Taylor expansion of Eq.~(\ref{eq:SNz}) for a flat universe around $z=0$:
\ba
(S/N)(z)\simeq \frac{A}{z}+\frac{1}{12} A (9 \Omega_{m,0}+8)-\frac{A}{144}\left(9 \Omega_{m,0} (9 \Omega_{m,0}-16)+16\right)z+\cdots,\ea
where $A$ is the proportionality constant in Eq.~(\ref{eq:SNz}) and $\Omega_{m,0}$ is the matter density today. Then, using the Lagrange inversion theorem, see Chapter 3.6.6. in Ref.~\cite{Abramowitz}, we invert the series in favor of the signal-to-noise $S/N$:
\ba
z&\simeq& \frac{A}{(S/N)}+\frac{A^2 (9 \Omega_{m,0}+8)}{12 (S/N)^2} +\frac{A^3 (6 \Omega_{m,0}+1)}{3 (S/N)^3} \cdots, \label{eq:seriesz}\ea
and we finally expand Eq.~(\ref{eq:dndsnz2}) around $z=0$ and use (\ref{eq:seriesz}) for the non-evolving GW sources, to find
\ba
\frac{dN}{d(S/N)}|_{flat}=\frac{4 \pi  A^4}{(S/N)^4}+\frac{32 \pi  A^5}{3 (S/N)^5}+\frac{10 \pi  A^6 (45 \Omega_{m,0}+70)}{9 (S/N)^6}+ \frac{4 \pi  A^7 (11 \Omega_{m,0}+4)}{(S/N)^7}+ \cdots . \nn \\ \label{eq:analz}
\ea
As can be seen in Eq.~(\ref{eq:analz}), the first term in the expansion goes as $\sim(S/N)^{-4}$ as expected for the low redshift limit, see Ref.\cite{Calabrese:2016bnu}. However, we do not expect this series expansion to be valid at redshifts higher than $z\simeq1$, as then the approximation breaks down. A similar expression can be found for Eq.~(\ref{eq:dndsnz3}), after assuming some functional form for the bias.

In Fig.~\ref{fig:series} we show the differential source counts $\frac{dN}{dx}$ as a function of the normalized signal-to-noise $x=\frac{S/N}{S/N_{min}}$ in the case of non-evolving GW sources for the numerical result (dashed line) and the series expansion of Eq.~(\ref{eq:analz}). For both cases we normalized $dN/dx$ to one at $x=1$, we used $\Omega_{m,0}=0.27$ and for the series expansion in addition we also matched its derivative at $x=4$ to that of the numerical case, so as to fix the value of $A$.

Finally, we also consider the case where the luminosity distance for a general dark energy model is given as a series expansion in terms of the {\it deceleration}, {\it jerk} and {\it snap} parameters defined as $q_0\equiv-\frac{1}{H_0^2}\frac{d^2a(t)}{dt^2}|_{t=t_0}$, $j_0\equiv\frac{1}{H_0^3}\frac{d^3a(t)}{dt^3}|_{t=t_0}$ and $s_0\equiv\frac{1}{H_0^4}\frac{d^4a(t)}{dt^4}|_{t=t_0}$ respectively. Then, we have that for a flat universe the luminosity distance is given by, see chapter 1.4 in Ref.~\cite{Weinberg:2008zzc}:
\ba
D_L(z)&=&\frac{c}{H_0}\left[z+\frac{1}{2} (1-q_0) z^2 \right. -\frac{1}{6} \left(1-q_0-3 q_0^2+j_0\right)z^3+\frac{1}{24}(2-2 q_0-15 q_0^2-15 q_0^3+5 j_0\nn \\&+&\left. 10 j_0 q_0+s_0)z^4+\cdots\vphantom{\frac{c}{H_0}}\right].\ea
Following the same procedure as before, and after using (\ref{eq:seriesz}) for the non-evolving GW sources, we find
\ba
\frac{dN}{d(S/N)}|_{flat}=\frac{4 \pi  A^4}{(S/N)^4}+\frac{32 \pi  A^5}{3 (S/N)^5}
+\frac{20 \pi  A^6 (3 q_0+10)}{9 (S/N)^6}
+\frac{8 \pi  A^7 (j_0+11 q_0+16)}{3 (S/N)^7}+\cdots .
\nn \\ \label{eq:analz1} \ea
A similar expression can be found for the case of the evolving GW sources , after assuming of course some functional form for the bias. Again, we see that the leading term is $\sim(S/N)^{-4}$ as expected.

To conclude, we have studied the intermediate redshift regime not only due to the complications of the full high redshift one, but also because the analytic approximations, e.g. Eqs.~(\ref{eq:analz})-(\ref{eq:analz1}) have proven to very interesting. These expressions provide insights on the behavior of the GW source counts but also the higher order corrections of the source counts in terms of the signal-to-noise $S/N$. For example, from Eqs.~(\ref{eq:analz})-(\ref{eq:analz1}) we can see the explicit dependence of the number counts $\frac{dN}{d(S/N)}$ on the cosmological parameters like $\Omega_{m,0}$ or $q_0$, something which is not possible to do in the high redshift case as it is a solely numerical approach, but also that the next leading order scales as $\sim(S/N)^{-5}$.

\subsubsection{Compactified extra dimensions}

In the section we will examine the effect of compactified extra dimensions on the source counts. For simplicity, given the abundance of possibilities, we will consider as an example a five-dimensional Kaluza-Klein (KK) theory, where the extra dimension has a compactification radius $R_c$. Then, if we assume that the metric does not depend on the extra dimension, much like in Kaluza's original theory, the five dimensional metric would be given by \cite{Bailin:1987jd}
\be
G^{(5)}_{AB}=\textrm{diag}\{g_{\mu\nu},-R_c^2\}, \label{eq:KKmetric}
\ee
where $g_{\mu\nu}$ is the usual four-dimensional metric. This will induce the existence of a KK graviton mode with equation, see for example Section 1.5 in Ref.~\cite{Bailin:1987jd}:
\be
\Box_5 h = \Box h + m_g^2 h =0\,.
\ee
The solution of the previous equation at large distances is a Yukawa type potential:
\be
h_{+\times} \propto \frac{e^{-R/R_c}}{R}\,.
\ee
Clearly, the exponential term acts as an effective cut-off on the flux of the BH binaries' emission of GW. In this case, we  have a similar result as in Ref.~\cite{Calabrese:2016bnu} but now the exponential cut-off is due to the effects of the compactified KK extra dimension. As before, the signal-to-noise is
\be
S/N \propto \frac{e^{-R/R_c}}{R} \label{eq:snR1}
\ee
or equivalently if we solve for $R$,
\be
R=R_c~\mathcal{W}\left(\frac{B}{R_c (S/N)}\right),
\ee
where $B$ is the constant of proportionality in Eq.~(\ref{eq:snR1}) and $\mathcal{W}(z)$ is the Lambert function\footnote{The Lambert function $\mathcal{W}(z)$ is the solution to the equation $f(z)=z e^z$ or $z=f^{-1}(z e^z)=\mathcal{W}(z e^z)$ and is implemented in Mathematica as ProductLog[z].}.

In this case, just like in the simple KK theory we only consider one extra dimension with size $R_c$. Therefore, the radial distribution of the source counts, since the gravitons also propagate on the extra dimension, will be given by:
\be
\frac{dN}{dR} \propto R^{2}R_c.\label{eq:dNdRKK}
\ee
Finally, the differential number counts will be given by:
\ba
\frac{dN}{d(S/N)}&=&\frac{dN}{dR}/\frac{d(S/N)}{dR} \nn \\
&=& -\frac{A R^4 R_c e^{R/R_c}}{B R+B R_c} \nn \\
&=& -\frac{A R_c^4 \mathcal{W}\left(\frac{B}{R_c (S/N)}\right)^3}{(S/N) \left(1+\mathcal{W}\left(\frac{B}{R_c (S/N)}\right)\right)}, \label{eq:dNdsnR1}
\ea
where in the last line we used Eq.~(\ref{eq:snR1}) to relate the distance $R$ to the signal-to-ratio and $A$ is the constant of proportionality in Eq.~(\ref{eq:dNdRKK}).

\begin{figure}[!t]
\centering
\vspace{0cm}\rotatebox{0}{\vspace{0cm}\hspace{0cm}\resizebox{0.80\textwidth}{!}{\includegraphics{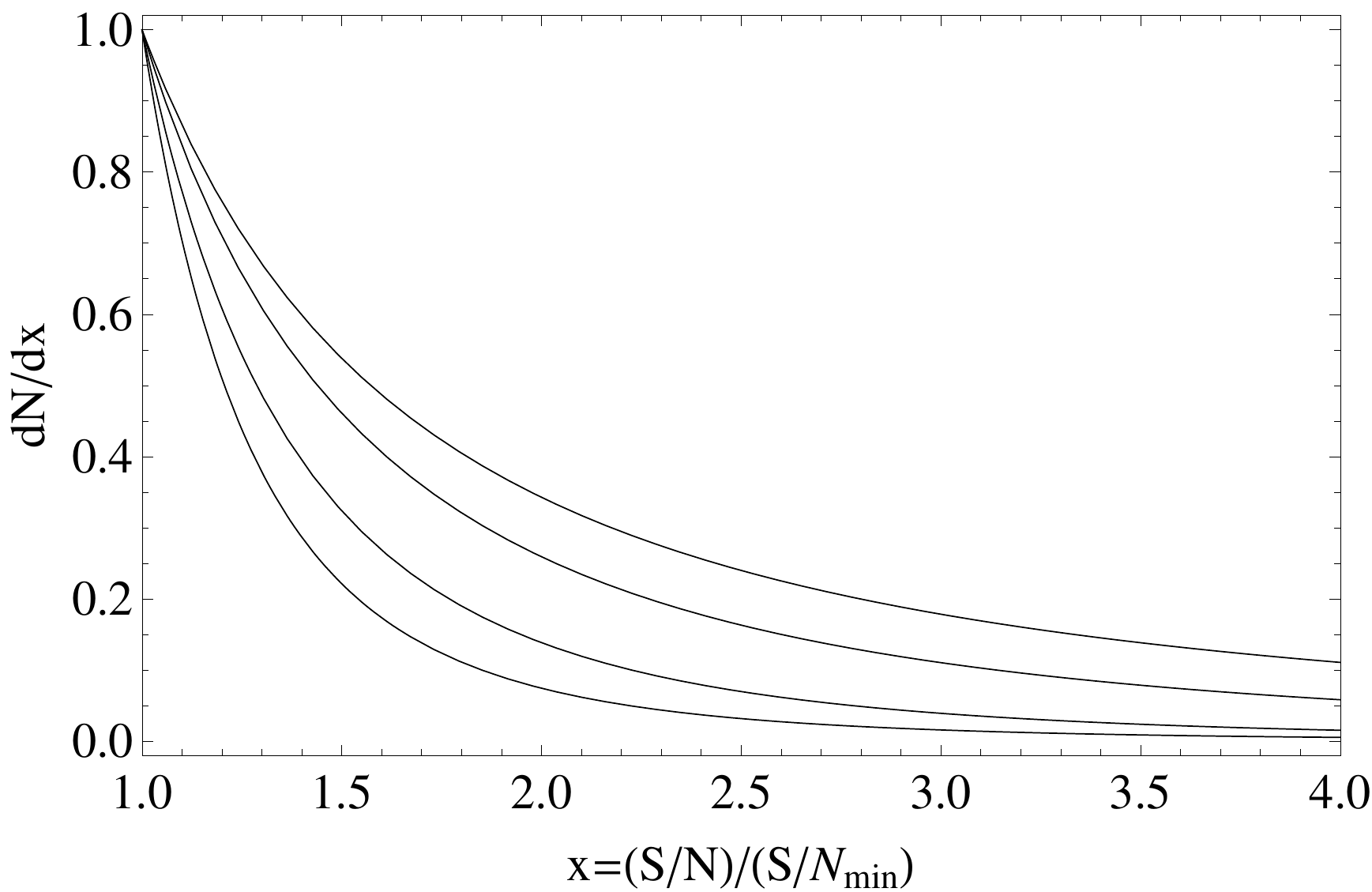}}}
\caption{The differential source counts $\frac{dN}{dx}$ as a function of the normalized signal-to-noise $x=\frac{S/N}{S/N_{min}}$. From bottom to top we have $\tilde{B}=(0.1,1,10,100)$. In all cases we have normalized the source counts to unity at $x=1$ for simplicity.  \label{fig:KK}}
\end{figure}

We can also renormalize the counts with respect to the minimum signal-to-noise $S/N_{min}$, so that if $S/N=x~S/N_{min}$, then we have:
\be
\frac{dN}{dx}=\frac{\tilde{A}~\mathcal{W}\left(\frac{\tilde{B}}{x}\right)^3}{x \left(1+\mathcal{W}\left(\frac{\tilde{B}}{x}\right)\right)},
\label{eq:dndxKK}
\ee
where $x\in[1,\infty]$, $\tilde{B}\equiv \frac{B/R_c}{ (S/N_{min})}$ and the constant $\tilde{A}\equiv -A~R_c^4/(S/N_{min})$ can be found by estimating the total number of events: $N_{tot}=\int_1^\infty \frac{dN}{dx} dx$ or $\tilde{A}=\frac{3~N_{tot}}{\mathcal{W}(\tilde{B})^3}$.

In Fig.~\ref{fig:KK} we show the theoretical curves for the differential source counts $\frac{dN}{dx}$ of Eq.~(\ref{eq:dndxKK}) as a function of the normalized signal-to-noise $x=\frac{S/N}{S/N_{min}}$. From bottom to top we have $\tilde{B}=(0.1,1,10,100)$. In all cases we have normalized the source counts to unity at $x=1$ for simplicity.

\section{Numerical results \label{sec:Numerics}}

In this section we present forecasts based on simulations of differential number counts detected at varying values of signal-to-noise $S/N$ above the current LIGO threshold $S/N_{min}=8$ for both cases studied before: the intermediate redshift limit and the compactified KK extra dimensions.

From Eqs.~(\ref{eq:analz}) and (\ref{eq:dndxKK}) for the intermediate redshift limit and compactified KK extra dimensions respectively, it follows that in all scenarios the differential number counts depend on two parameters. These are $A$ and $\Omega_{m,0}$ for the first case and $\tilde{A}$ and $\tilde{B}$ for the second. The aim of this section is to provide forecasted constraints in each of these parameters.

In general, there are are two ways one can can define the likelihood and a \textit{chi-square} statistic $\chi^2$ in order to extract the model parameters. The first would be to define the log-likelihood describing the sample of data points as \cite{Cash:1979vz}
\ba
\ln \mathcal{L} & = & \ln \prod_{i=1}^{N_{b}} P\bigg(\frac{dN}{dx}_i\bigg)  \\
       & = & \sum_{i=1}^{N_{b}} \bigg(d_i \ln \frac{dN}{dx}_i-\frac{dN}{dx}_i\bigg).
\ea
The other method, which is also the one that we consider here, is to define the \textit{chi-square} as usual:
\ba
\chi^2&=&\sum_{i=1}^{N_{tot}}\left(O_i-E_i\right)^2,
\label{eq:chi2}
\ea
where $O_i$ is the number of events and $E_i=\frac{dN}{dx}_i$ is the expected number counts, both at a particular $x_i$. It should be noted that it would be more proper to use the Poisson distribution for small samples and the Gaussian distribution for larger ones. However, we have chosen to use the latter as for several tests we did, we found very small and almost negligible differences between the results recovered by both methods.

We now simulate a random realization of 10, 50, 100 and 500 sample sources and calculate the likelihood for each of these realizations. These number of detections are realistic based on the pessimistic high BH kicks model (few detections per year), the standard model rate with rates of about 200 detections per year (LIGO single detector) and 500 $\textrm{yr}^{-1}$ for a 3-detector network and the optimistic model with 1000-3000 detections $\textrm{yr}^{-1}$, as seen in Table 1 of Ref.~\cite{Belczynski:2016obo} but see also Ref.~\cite{Belczynski:2010tb}. There are two approaches to build these samples.

A first approach would be to distribute sources homogeneously over $S/N$ up to a maximum threshold, here taken at $S/N_{max} = 32$ or $x = 4$. This method obviously contradicts the fact that the number counts decrease with increasing $S/N$. A second, more realistic approach, is to place the sources with increased probability at low $S/N$. In this case, the probability distribution function used to generate the sources at $(x,x+dx)$ is given by:
\be
P(x) = D x^{-4}, \quad x\in[x_{max}^{-1/4},x_{min}^{-1/4}],
\label{eq:prob_x}
\ee
where $x_{min} = 1$, $x_{max} = 4$ are the normalized $S/N$ of the threshold and strongest event, and $D$ is a normalization constant.

We then create mock realizations for the simulated differential number counts $dN/dx$ corresponding to each $x$ and we can calculate the $\chi^2$ by using Eq.~(\ref{eq:chi2}), and thus the likelihood. As mentioned earlier, we have checked that our choices for the data analysis are robust and are in agreement at the appropriate limits with Ref.~\cite{Calabrese:2016bnu}. With the likelihood finally available, we are at last able to perform a MCMC sampling of the posterior\footnote{The MCMC analysis was performed with codes developed by the authors. One such generic MCMC sampler made by one of the authors can be found at \href{www.uam.es/savvas.nesseris/codes.html}{www.uam.es/savvas.nesseris/codes.html}}. The following subsections deal with the prior choices, chain specifications and constraints on the parameters, for the three different cases with each with 10, 50, 100 and 500 detected sources.

\subsection{Intermediate redshift limit}

We perform the MCMC analysis in the case of the intermediate redshift limit with the theoretical prediction for $\frac{dN}{dx}$ given by Eq.~(\ref{eq:analz}), which has two free parameters $A$ and
$\Omega_{m,0}$ for a flat universe, and assuming a $\Lambda$CDM model with $\Omega_{m,0}=0.27$. The original input parameters for $\tilde{A}$ were chosen so that Eq.~(\ref{eq:analz1}) integrated over all $S/N$ gives the total number of events. Specifically, for 10, 50, 100 and 500 events we find $\tilde{A}=$4.2043, 5.4868, 6.1012 and 7.6840 respectively.

\begin{figure*}[!t]
\centering
\vspace{0cm}\rotatebox{0}{\vspace{0cm}\hspace{0cm}\resizebox{0.48\textwidth}{!}{\includegraphics{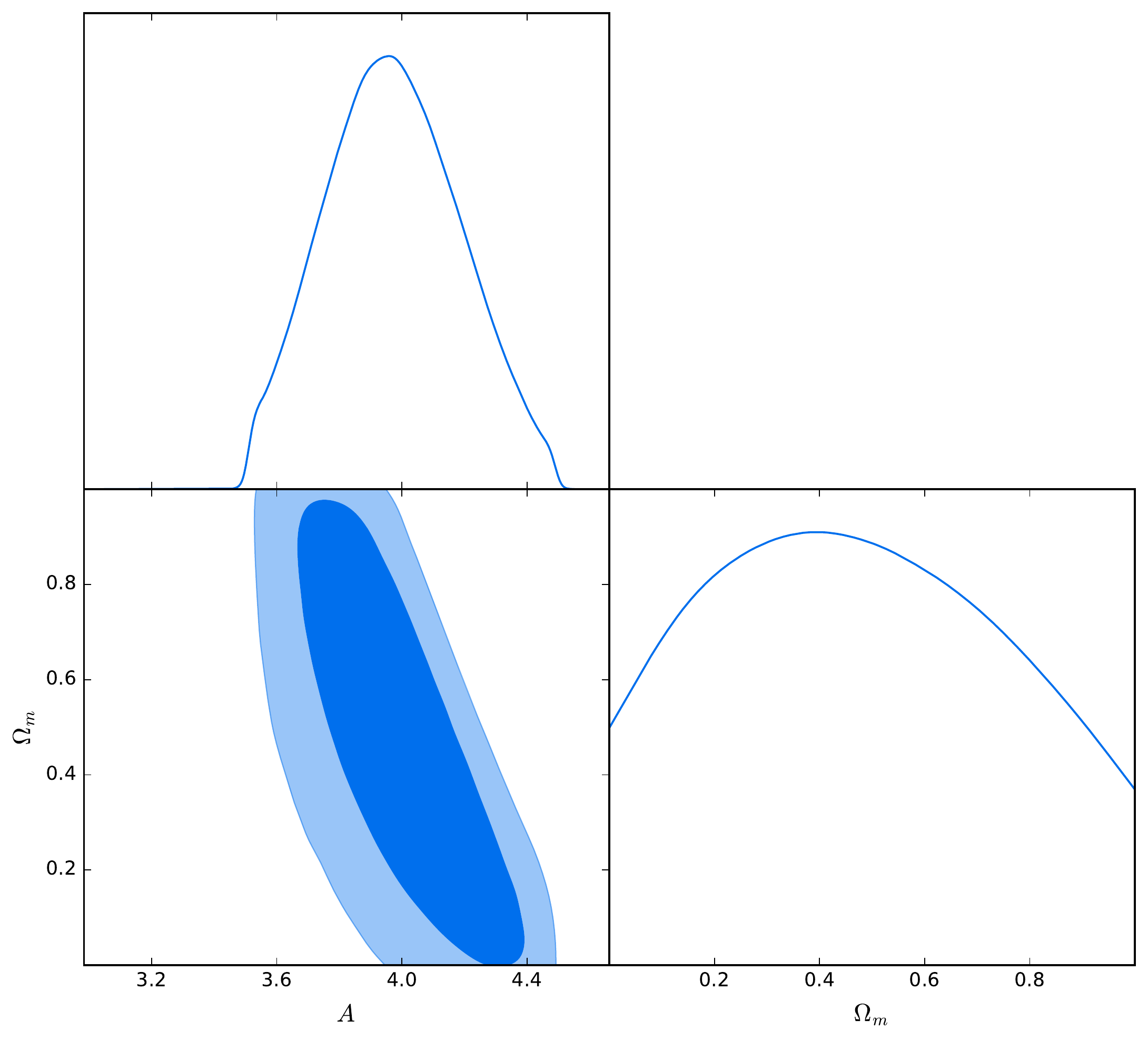}}}
\vspace{0cm}\rotatebox{0}{\vspace{0cm}\hspace{0cm}\resizebox{0.48\textwidth}{!}{\includegraphics{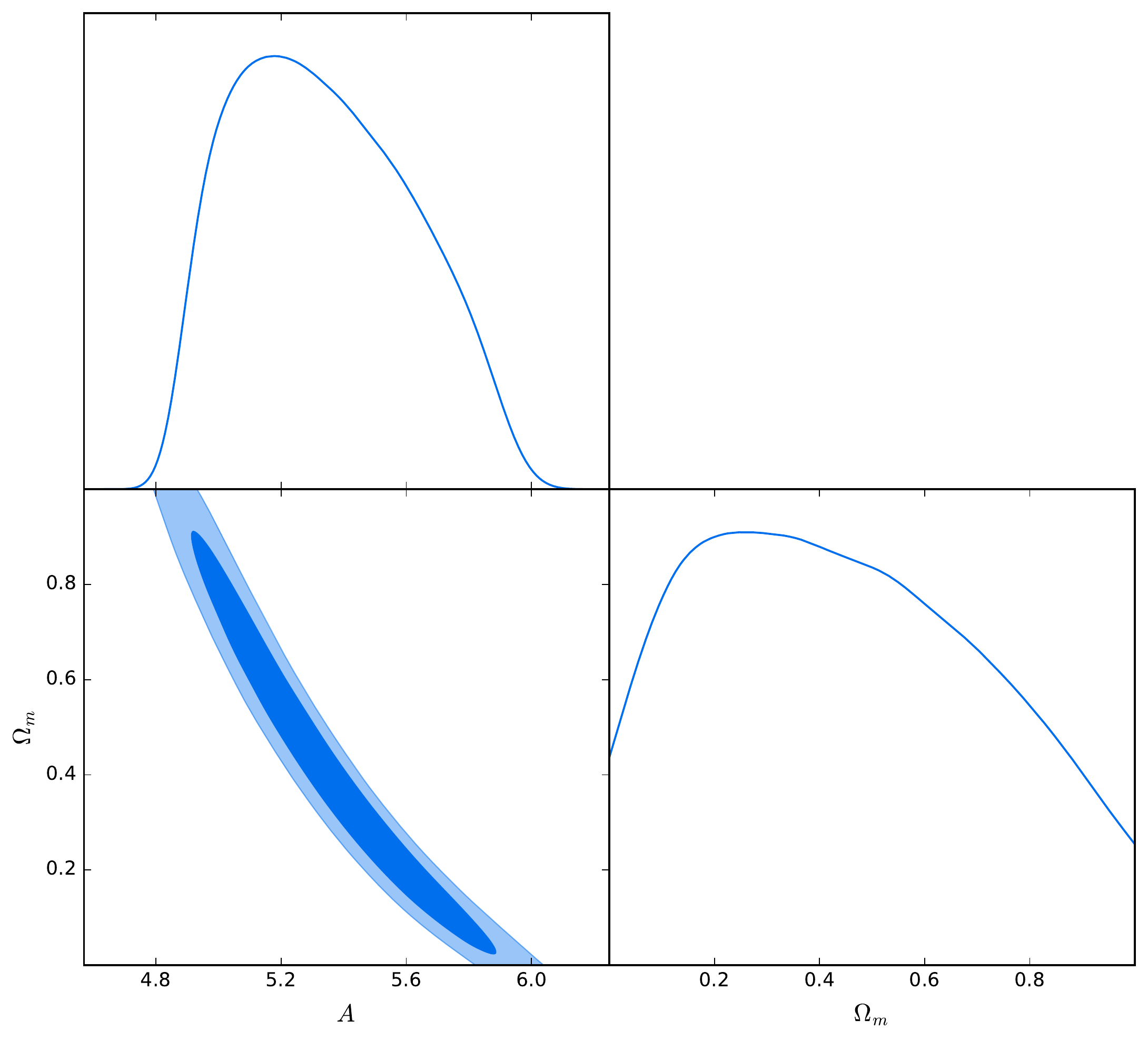}}}
\vspace{0cm}\rotatebox{0}{\vspace{0cm}\hspace{0cm}\resizebox{0.48\textwidth}{!}{\includegraphics{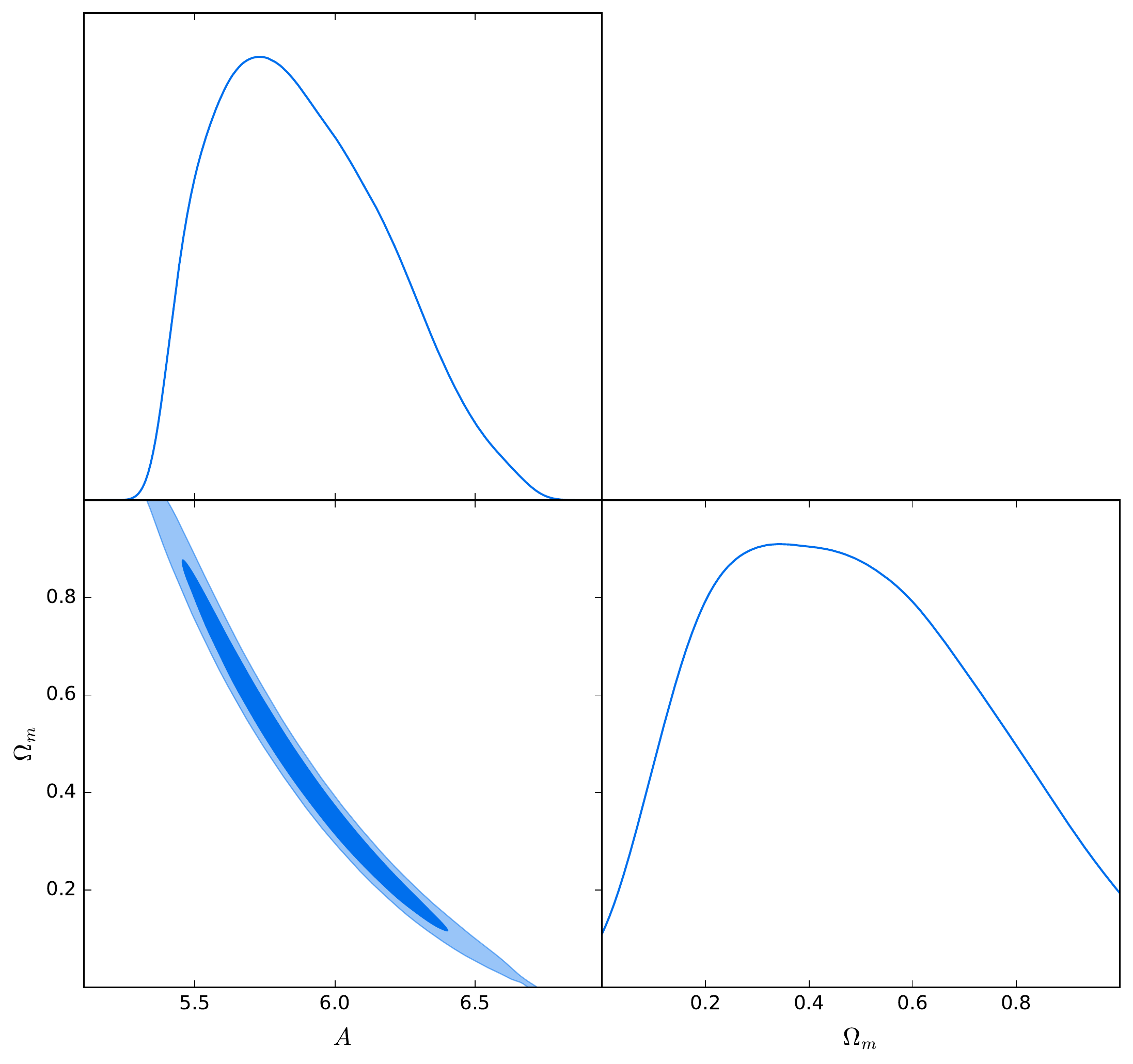}}}
\vspace{0cm}\rotatebox{0}{\vspace{0cm}\hspace{0cm}\resizebox{0.48\textwidth}{!}{\includegraphics{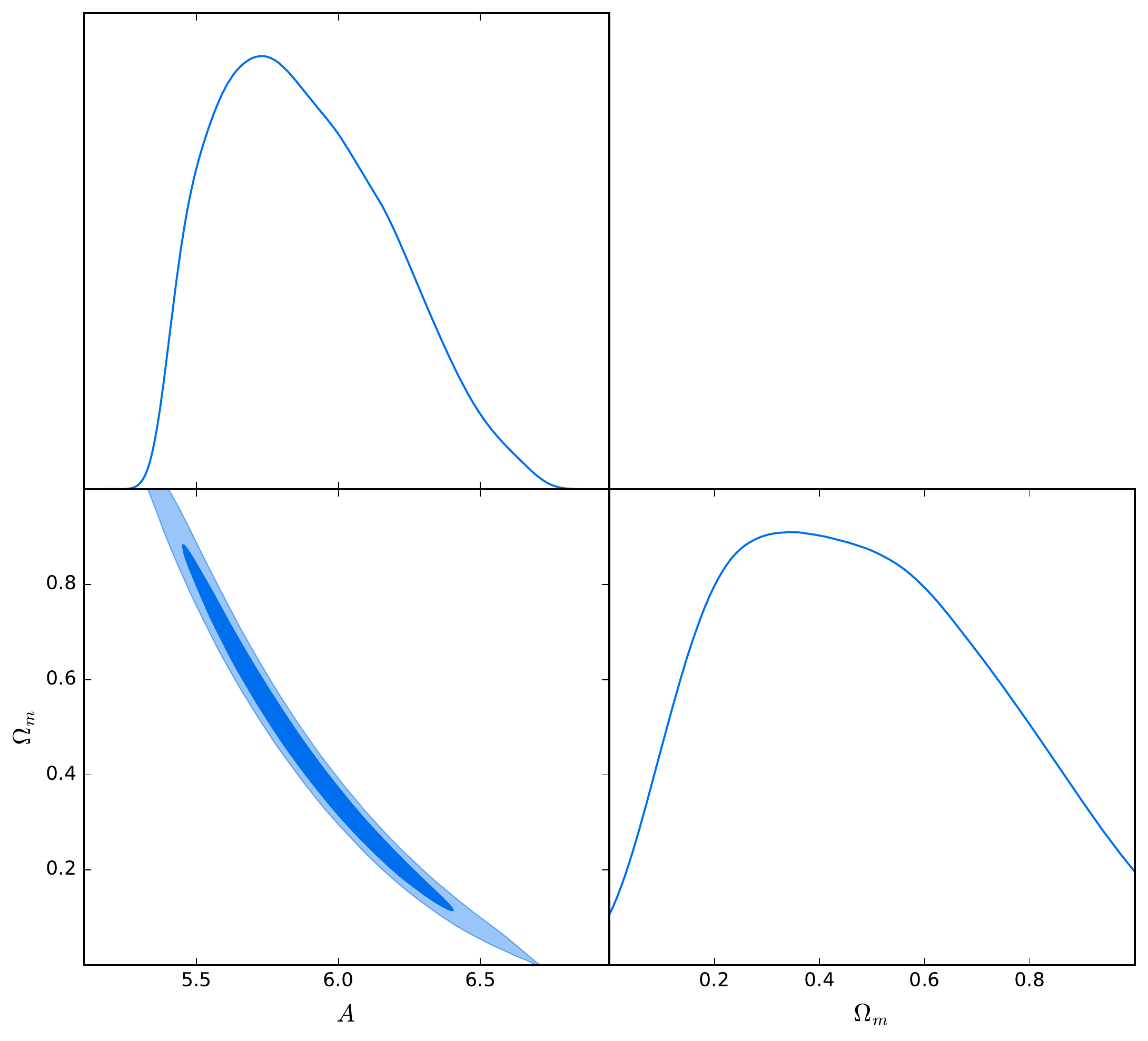}}}
\caption{The 2D $68.3\%$ and $95.5\%$ confidence contours and the 1D marginalized posteriors for the case of the intermediate redshift limit given by Eq.~(\ref{eq:analz}) for the two free parameters $A$ and
$\Omega_{m,0}$. The plots correspond to 10 (top left), 50 (top right), 100 (bottom left) and 500 (bottom right) events respectively. The original input parameters for the mock data are $\Omega_{m,0}=0.27$ and $\tilde{A}$ chosen so that Eq.(\ref{eq:analz1}) integrated over all $S/N$ gives the total number of events as mentioned in the text. For 10, 50, 100 and 500 events we find $\tilde{A}=$4.2043, 5.4868, 6.1012 and 7.6840 respectively. \label{fig:highz_plots}}
\end{figure*}

In Fig. \ref{fig:highz_plots} we show the 2D $68.3\%$ and $95.5\%$ confidence contours and the 1D marginalized posteriors for the case of the intermediate redshift limit given by Eq.~(\ref{eq:analz}) for the two free parameters $A$ and $\Omega_{m,0}$. The plots correspond to 10 (top left), 50 (top right), 100 (bottom left) and 500 (bottom right) events respectively. As can be seen, with a small number of events it is not possible to constrain the geometry of the universe, hence $\Omega_{m,0}$, but the situation improves dramatically with 500 events when the best-fit value for the matter density is $\Omega_{m,0}=0.280\pm0.083$, perfectly consistent with the real value of $\Omega_{m,0}=0.27$. A total of 500 events could be readily achievable with future detectors \cite{Belczynski:2010tb,Belczynski:2016obo}.\\

\begin{figure*}[!t]
\centering
\vspace{0cm}\rotatebox{0}{\vspace{0cm}\hspace{0cm}\resizebox{0.48\textwidth}{!}{\includegraphics{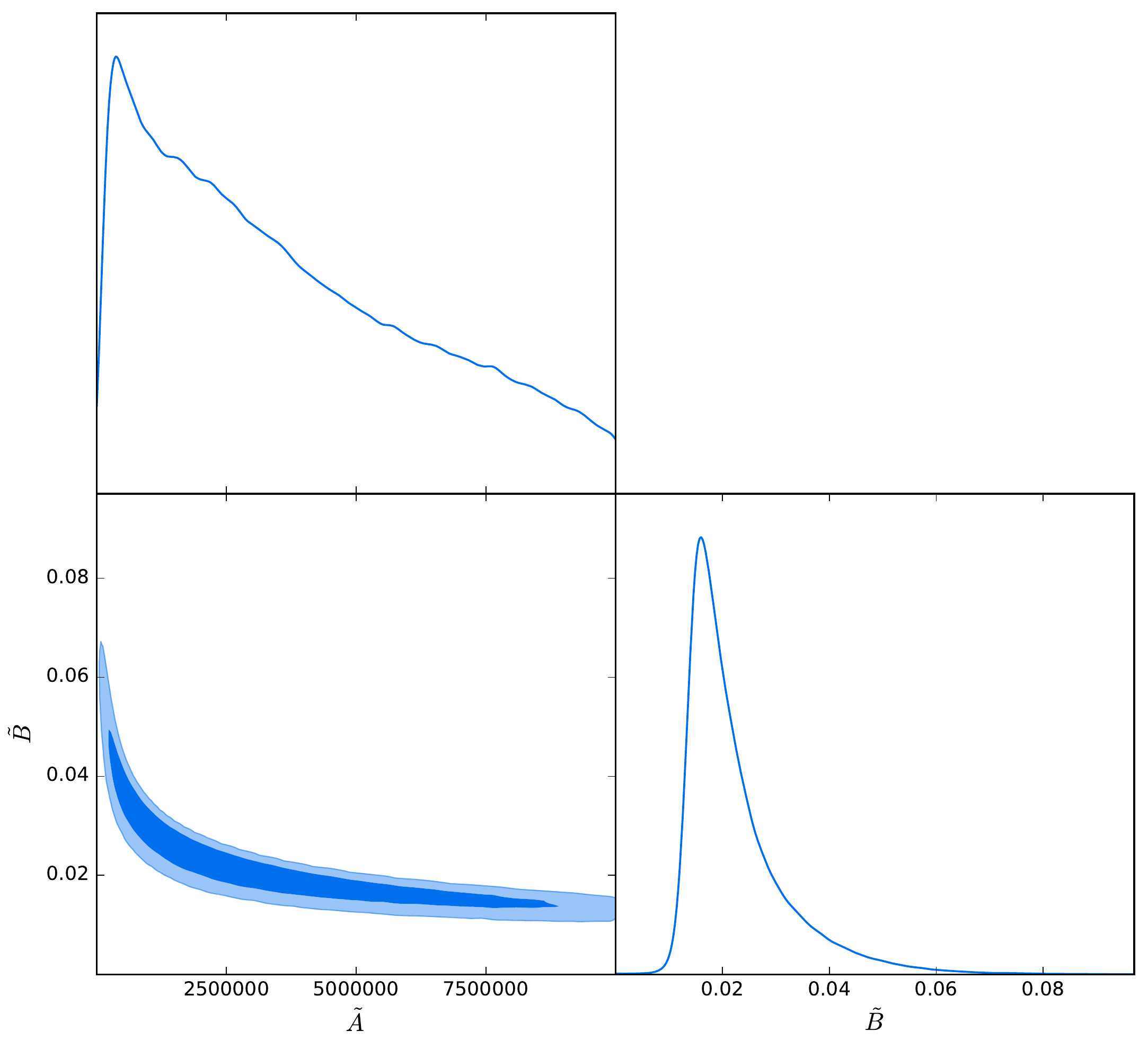}}}
\vspace{0cm}\rotatebox{0}{\vspace{0cm}\hspace{0cm}\resizebox{0.48\textwidth}{!}{\includegraphics{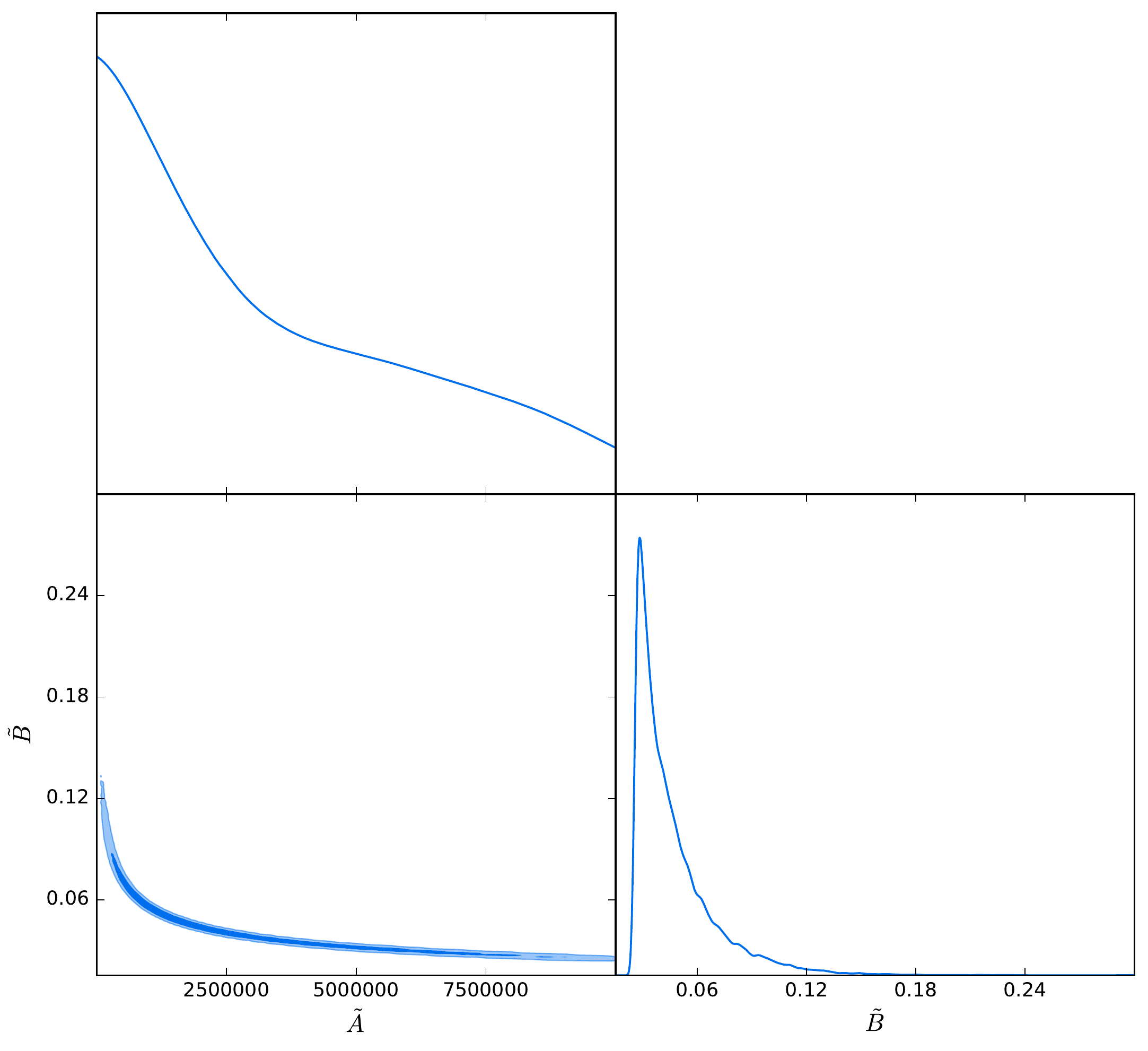}}}
\vspace{0cm}\rotatebox{0}{\vspace{0cm}\hspace{0cm}\resizebox{0.48\textwidth}{!}{\includegraphics{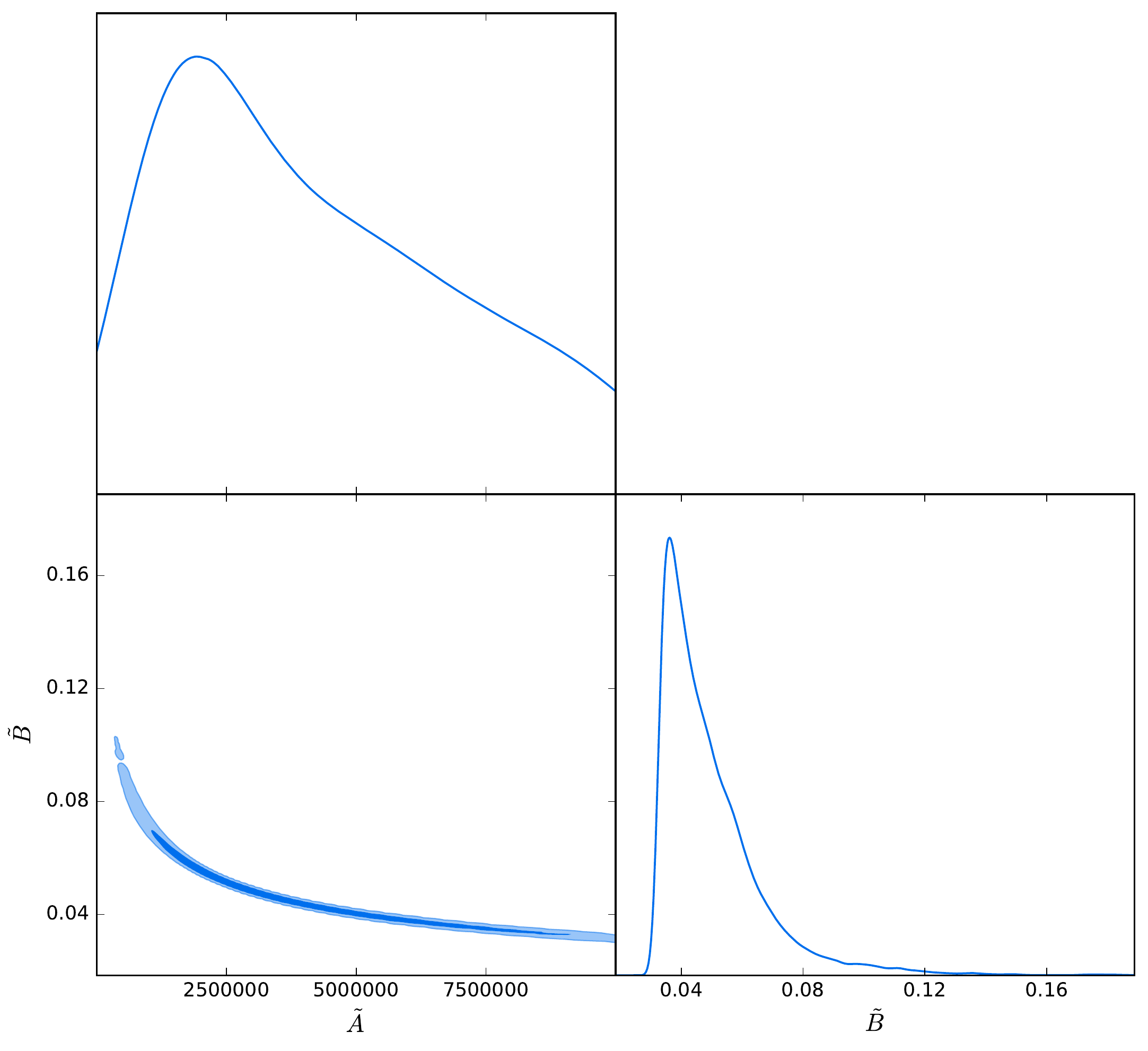}}}
\vspace{0cm}\rotatebox{0}{\vspace{0cm}\hspace{0cm}\resizebox{0.48\textwidth}{!}{\includegraphics{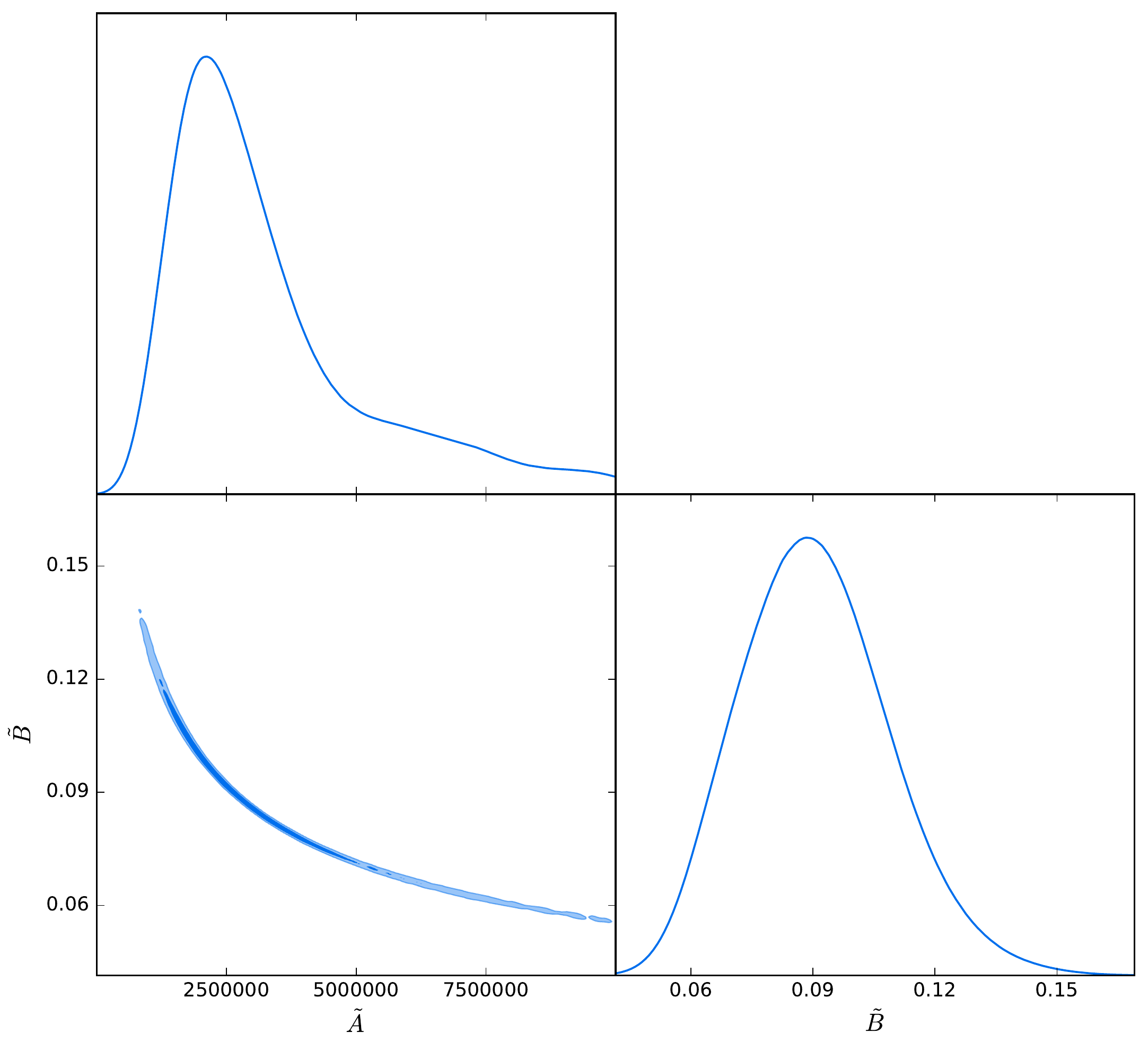}}}
\caption{The 2D $68.3\%$ and $95.5\%$ confidence contours and the 1D marginalized posteriors for the case of the compactified KK extra dimension given by Eq.~(\ref{eq:dndxKK}) for the two free parameters $\tilde{A}$ and $\tilde{B}$. The plots correspond to 10 (top left), 50 (top right), 100 (bottom left) and 500 (bottom right) events respectively. The original input parameters for the mock data are $\tilde{B}=0.1$ and $\tilde{A}= 3~N_{tot}/\mathcal{W}(\tilde{B})^3$ as mentioned in the text. \label{fig:Compactified_rest}}
\end{figure*}

\subsection{Compactified extra dimensions}

In this case we perform the MCMC analysis in the case of the compactified extra dimensions with the theoretical prediction for $\frac{dN}{dx}$ given by Eq.~(\ref{eq:dndxKK}). Note that the free parameters are $\tilde{B}$ and $\tilde{A}$ and are related through the number of observed events, by $\tilde{A}= 3~N_{tot}/\mathcal{W}(\tilde{B})^3$. It is then expected that for a low number of sources, they will be degenerate with respect to each other. In contrast with the previous case, we have not found that a larger number of events may break the degeneracy between the two.

For all four scenarios with $N_{obj}=10, 50 ,100, 500$ we take the limiting case of infinite compactification radius for which $\tilde{B}=0$ as the lower bound of the prior. We find, however, that no upper bound is able to capture the full confidence region, which is to be expected if the degeneracy between the two parameters remains unbroken for a larger number of events. Moreover, as mentioned before, the parameter $B$ is the constant of proportionality in Eq.~(\ref{eq:snR1}), while the parameter $\tilde{B}$ is related to $B$, the radius of compactification $R_c$ and the minimum signal-to-noise $S/N_{min}$. Giving a priori an expectation value for the parameter $B$ is not possible, however using physical intuition and expecting that the number counts will be non-zero, one can assume a prior region $\tilde{B}\in[0,1]$, and thus we choose $\tilde{B} \in [0,1]$ and $\tilde{A} \in [0,10^7]$. The original input parameters for the mock data are $\tilde{B}=0.1$ and $\tilde{A}= 3~N_{tot}/\mathcal{W}(\tilde{B})^3$.

Again we map out the posterior function and find constraints on the parameters by running the MCMC up to $10^5$ samples in each of the scenarios considered. Convergence is quickly achieved and the burn-in phase barely takes more than $10^3$ steps before the chains becomes Markovian. In this case the 2D $68.3\%$ and $95.5\%$ confidence contours and the 1D marginalized posteriors for the two free parameters $\tilde{A}$ and $\tilde{B}$ can be seen in Fig.~\ref{fig:Compactified_rest}. As can be seen in the plots, in this case the model parameters are highly degenerate with respect to each other even for 500 events. Furthermore, it can be seen that the range for the 1-D distributions for $\tilde{B}$ tends to increase. The reason for this is that this parameter depends implicitly and proportionally on the number of events, thus tends to increase with increasing numbers of events.


\section{Discussion and Conclusions}

In this paper we generalized the source counts $\frac{dN}{d(S/N)}$ methodology to all redshifts. This was possible under certain assumptions like knowledge of the redshift-dependent BH to galaxy bias $b_{BH}(z)$, the underlying cosmology and a source model for the black hole binary mergers. First, we found that for a given configuration of the BH-BH binary system that produces the GW event in the high redshift regime one would potentially expect two windows where observations above the minimum signal-to-noise threshold can be made. This holds assuming there are no higher order corrections in the redshift dependence of the signal-to-noise ratio $S/N(z)$ for the expected prediction. However, it should be noted that in general the GW emission and the observed strain will also depend on the properties of the sources and the chirp masses. So, for example, an object at a specific redshift $z$ can be seen at a signal-to-noise which could be substantially different if the masses of the BHs and the inclination of the orbits are different.

On the other hand, the expression for the signal-to-noise of Eq.~(\ref{eq:SNz}) is valid only for small to intermediate redshifts, thus making the minimum of the signal-to-noise potentially artificial. In this regard, we proposed a regularization method that solves the issue of the singularity in the number counts in the high redshift regime. We have left the determination of the possible higher order redshift corrections to  Eq.~(\ref{eq:SNz}) for future work.

Another important aspect is whether the GW sources follow the distribution of their hosts or the less realistic case where they simply evolve with the volume $N_{BH}\sim (1+z)^3$. We took both possible effects into account and specifically in the first case we allowed for a BH to galaxy bias $b_{BH}(z)$, as suggested by Ref.~\cite{Oguri:2016dgk}, and we also included the proper galaxy redshift distribution as a function of the median redshift of the survey $z_0$. In general, taking all these effects into account means that calculating the source counts $\frac{dN}{d(S/N)}$ is possible only numerically and requires certain assumptions, such as the knowledge of the BH to galaxy bias $b_{BH}(z)$.

However, in the simple case where we can assume the GW sources simply evolve with the volume $N_{BH}\sim (1+z)^3$ it is possible to calculate the source counts, as we had done in Eq.~(\ref{eq:analz}), by a series approximation. This can done at intermediate redshifts $z\lesssim1$ for the cosmological constant model or by Eq.~(\ref{eq:analz1}) for any dark energy model described by the cosmographic parameters $(q_0,j_0,s_0)$ in a flat universe. The source counts for the general case and for all redshifts, unfortunately cannot be calculated analytically.

After implementing an MCMC analysis with a varying number of events, 10, 50, 100 and 500 respectively, we found that for a small number of events the number counts are completely insensitive to the value of the matter density $\Omega_{m,0}$ of the flat cosmological constant model. On the contrary, for a high enough number, i.e. 500, we find a constraint of $\Omega_{m,0}=0.280\pm0.083$, perfectly consistent with the real value of $\Omega_{m,0}=0.27$ used for the mock data. This constraint might not be competitive to the level of the of Planck at the moment, but this methodology could be used to break degeneracies with other data and also to tightly constrain the relation between the absolute distance and the redshift as proposed in Ref.~\cite{Oguri:2016dgk}. Moreover, these constraints are expected to improved in the near future with the expected increase in the number of events but also better understanding of the systematics and the underlying physics of the BH binary mergers.

We also considered a case with extra dimensions in the low redshift limit $(z\ll1)$, specifically a Kaluza-Klein model with an extra dimension with a compactification radius $R_c$. After doing the MCMC forecasts as before, we found that in the case of the KK extra dimension the degeneracies are not fully broken, but instead the parametric space is just made smaller. In any case, the methodology as described remains promising as a detection of a high number of events even at high redshifts $z$ should be possible with future detectors.


\begin{acknowledgments}
The authors would like to thank C.~Belczynski for useful discussions and an anonymous referee for useful and constructive comments. This work is supported by the Research Project of the Spanish MINECO, FPA2013-47986-03-3P, and the Centro de Excelencia Severo Ochoa Program SEV-2012-0249. S.~N. is supported by the Ram\'{o}n y Cajal programme through the grant RYC-2014-15843.

\end{acknowledgments}

\bibliographystyle{utcaps}
\bibliography{gravity_waves_jcap}

\end{document}